\documentclass[aps,prl,twocolumn,superscriptaddress,showpacs,bibnotes]{revtex4-1}
\usepackage{amsmath}

\usepackage{amsfonts}
\usepackage{soul}
\usepackage{xcolor}
\usepackage{amssymb}
\usepackage{graphicx}
\usepackage{hyperref}
\usepackage[all]{hypcap}

\hypersetup{backref,
	pdfpagemode=FullScreen,
	colorlinks=true, citecolor=blue}
\usepackage[sort&compress]{natbib}
\begin{document}
\title{A phononic crystal coupled to a transmission line via an artificial atom} 
\author{Aleksey N. Bolgar}
\email[]{alexgood@list.ru; the corresponding author}
\affiliation{Laboratory of Artificial Quantum Systems, Moscow Institute of Physics and Technology, 141700 Dolgoprudny, Russia}

\author{Daniil D. Kirichenko}
\email[]{dkd1274@ya.ru}
\affiliation{Laboratory of Artificial Quantum Systems, Moscow Institute of Physics and Technology, 141700 Dolgoprudny, Russia}

\author{Rais. S. Shaikhaidarov}
\email[]{R.Shaikhaidarov@rhul.ac.uk}
\affiliation{Physics Department, Royal Holloway, University of London, Egham, Surrey TW20 0EX, United Kingdom}
\affiliation{Laboratory of Artificial Quantum Systems, Moscow Institute of Physics and Technology, 141700 Dolgoprudny, Russia}

\author{Shtefan V. Sanduleanu}
\email[]{shtefan.sanduleanu@gmail.com}
\affiliation{Laboratory of Artificial Quantum Systems, Moscow Institute of Physics and Technology, 141700 Dolgoprudny, Russia}

\author{Alexander V. Semenov}
\email[]{a_sem2@mail.ru}
\affiliation{Laboratory of Artificial Quantum Systems, Moscow Institute of Physics and Technology, 141700 Dolgoprudny, Russia}
\affiliation{Moscow State Pedagogical University, 119991 Moscow, Russia}

\author{Aleksey Yu. Dmitriev}
\email[]{dmitrmipt@gmail.com}
\affiliation{Laboratory of Artificial Quantum Systems, Moscow Institute of Physics and Technology, 141700 Dolgoprudny, Russia}

\author{Oleg V. Astafiev}
\email[]{Oleg.Astafiev@rhul.ac.uk; the corresponding author}
\affiliation{Skolkovo Institute of Science and Technology, 121205 Moscow, Russia}
\affiliation{Laboratory of Artificial Quantum Systems, Moscow Institute of Physics and Technology, 141700 Dolgoprudny, Russia}
\affiliation{Physics Department, Royal Holloway, University of London, Egham, Surrey TW20 0EX, United Kingdom}
\affiliation{National Physical Laboratory, Teddington, TW11 0LW, United Kingdom}
\date{\today}

\begin{abstract} 
We study a phononic crystal interacting with an artificial atom -- a superconducting quantum system -- in the quantum regime. The phononic crystal is made of a long lattice of narrow metallic stripes on a quatz surface. The artificial atom in turn interacts with a transmission line therefore two degrees of freedom of different nature, acoustic and electromagnetic, are coupled with a single quantum object. A scattering spectrum of propagating electromagnetic waves on the artificial atom visualizes acoustic modes of the phononic crystal. We simulate the system and found quasinormal modes of our phononic crystal and their properties. The calculations are consistent with the experimentally found modes, which are fitted to the dispersion branches of the phononic crystal near the first Brillouin zone edge. Our geometry allows to realize effects of quantum acoustics on a simple and compact phononic crystal. 
\end{abstract}

\maketitle

{\bf Introduction}

Superconducting\cite{wallraff2004strong} quantum systems are promising for prospective quantum technologies, particularly for quantum informatics. Such systems are also very interesting for fundamental physics, especially for quantum optics with artificial atoms \cite{wallraff2004strong,koch2007charge,Mooij2003flux,Martinis2002} and for implementing them in new research directions. Recently, several works have founded Quantum Acoustodynamics (QAD) with the artificial atoms, where electromagnetic waves are replaced by acoustic ones and photons by phonons. \cite{POOT,Aspelmeyer,Treutlein}

One of the key elements in the QAD experiments is a mechanical resonator, which can be a bulk resonator or a surface acoustic wave resonator, playing the similar role as a cavity in Quantum Electrodynamics (QED). Since the wavelength of acoustic waves is typically five orders of magnitude shorter than that of electromagnetic waves, acoustic elements can be made much more compact. Pioneering experiments in the area were made with bulk acoustic resonators coupled to superconducting qubits.  \cite{o2010quantum,chu2017quantum} However, integration of such bulk resonators with electronics is not straightforward.

On the other hand, surface acoustic wave (SAW) resonators can be directly combined with superconducting quantum circuits because they are shaped as planar metallic structures on the surface of piezoelectric. However, it is a technologically challenging problem to reach the quantum regime for the SAW resonators, since it requires state-of-the-art nano-fabrication methods. A series of experiments have already demonstrated superconducting qubits interacting with SAW. The experiment in Ref.~[\citenum{kockum2014designing}] demonstrated interaction of SAW with a qubit of the transmon type. Later, SAW resonators working in the gigahertz range (required for quantum regime) were shown. They have quality factors sufficiently high for use in qubit readout schemes. \cite{manenti2016surface} Recently, our group has demonstrated the quantum regime of a SAW resonator. \cite{bolgar2018quantum} In that work, we observed a Rabi-splitting caused by an artificial atom coupled to resonator modes. Similar methods with electromagnetic resonators are well known from quantum electrodynamics (QED) experiments. \cite{koch2007charge,wang2008measurement}

\begin{figure}[h] 
	
	\includegraphics[width=1.0\columnwidth]{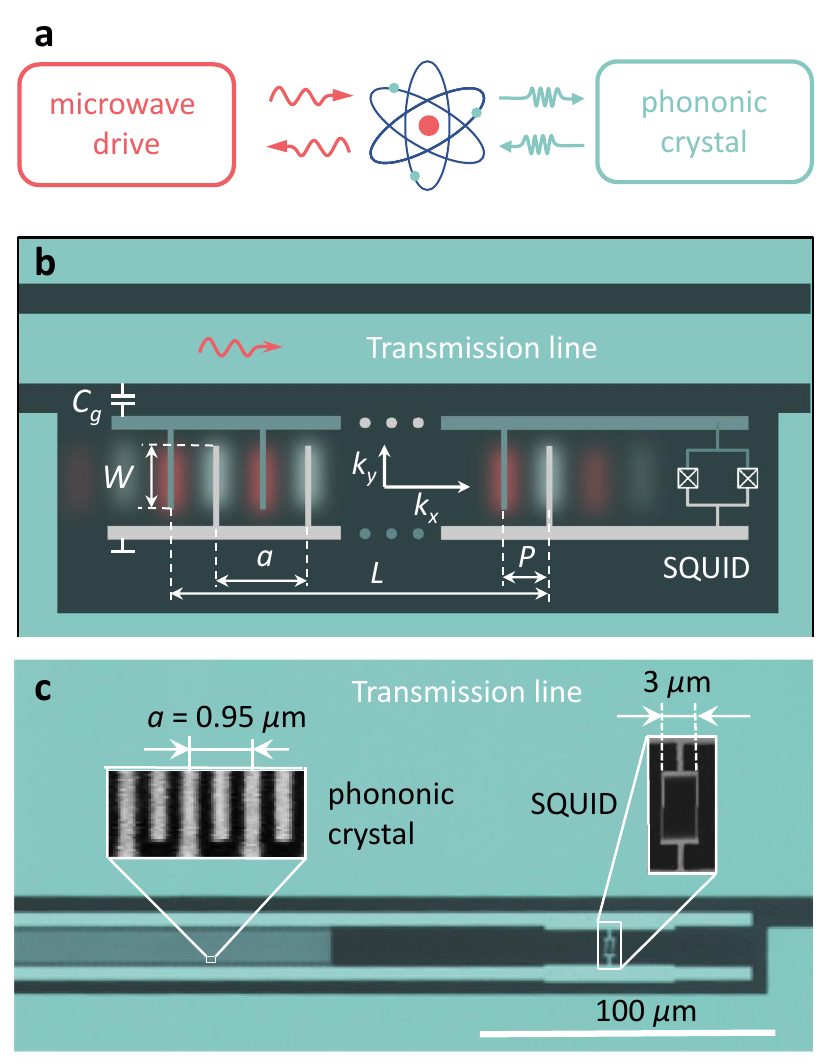}
	\caption{{\bf{The device.}} {\bf{a}}, Schematics of the device. The artificial atom is simultaneously coupled to electromagnetic and acoustic systems. Microwave photons excite an artificial atom (qubit). The atom in turn generates phonons into the phononic crystal.  
	{\bf{b}}, Schematic representation of the sample. Electromagnetic waves propagate through a coplanar transmission line  and interact with an artificial atom shaped as a transmon. The qubit shunting capacitance consists of $N_p=140$ identical electrode pairs (metallic stripes). The corresponding mechanical substrate surface oscillations are shown by color gradients.
	{\bf{c}}, Micrograph of the sample. Thin structures of the phononic crystal and the SQUID are shown in the insets. 
	\label{Fig1} 
	}
	
\end{figure}

In this work, we study a hybrid circuit -- an artificial atom strongly coupled simultaneously to two systems of different nature (acoustic and electromagnetic) -- a phononic crystal and a 1D transmission line of electromagnetic waves. Here we utilize a unique property of superconducting quantum systems: they can easily achieve the strong coupling regime to macroscopic circuit elements; strong coupling can be achieved to several of the circuit elements. \cite{Peng2018aulterThomas} In other words, our quantum system is intermediate between the phononic crystal and a coplanar transmission line as shown in Fig.~\ref{Fig1}(a,b). In contract to the previous experiments, our artificial atom is coupled to a single long phononic crystal without additional elements (e.g. mirrors). In spite of simple geometry, the system can be used to demonstrate a variety of effects of Quantum Acoustics. It scales down area of the acoustic part of the device at least an order of magnitude, making it more compact and making the fabrication process of the periodic structure much easier and more robust. Thus, the system with a phononic crystal has a significant technological advantage.  

The phononic crystal is physically different from systems with resonators studied, for example, in Ref.~[\citenum{bolgar2018quantum}]. In an ideal resonator, a field is confined and quantized. Differently, in an open phononic crystal the radiation is freely leaking away from the boundaries and the quasinormal modes, describing acoustic fields, are not quantized. In real systems, dissipation in resonators must be accounted for. On the other hand, the allowed modes of long phononic crystals can be approximated by quantized modes. In Ref.~[\citenum{bolgar2018quantum}], a quantum system has been coupled to modes of a resonator. In this work, we study an open phononic crystal formed by an open system of an array of metallic stripes on a quartz surface.

We study a spectrum of the hybrid system by measuring scattered electromagnetic waves. In the spectrum, we found coherent resonant interactions of the atom with several modes of the phononic crystal, which had the strongest coupling to the atom. Similar system has been analysed theoretically before. \cite{Delsing2018cavityfree} The modes frequencies are on the branches of the dispersion curves of the crystal near the edge of the first Brillouin zone. 
\\

{\bf The device layout }

Our device and its operation principals are shown in Fig.~\ref{Fig1}. It is fabricated on a piezoelectric substrate of ST-x cut of quartz. The device consists of a transmon-type qubit \cite{koch2007charge} capacitively coupled to a microwave transmission line. The transmon shunting capacitance has a form of an interdigital transducer (IDT) with 30 nm thick and about 250 nm wide equally spaced electrodes in the form of metallic stripes of $W$~=~12~$\mu$m length. The IDT capacitance ($C_q\approx 83$ fF) is proportional to the number of the electrode pairs $N_p$~=~140. With the IDT period $a\approx 0.95~\mu$m, the generation of surface acoustic waves (SAWs) is the most efficient at frequencies close to $f_{ac} = v/a\approx$~3.3~GHz, where $v=3.16$ km/s is the speed of SAW for quartz and the total length of the structure is $L = N_p a$~=~133~$\mu$m. The capacitance electrodes are connected to a SQUID loop to tune the qubit energies by an external magnetic field. The sizes of the SQUID Josephson junctions are 100$\times$100 nm$^2$ and its maximal Josephson energy $E_J/h =$~9.6~GHz. 
The transmon is additionally coupled to a transmission line with the capacitance $C_g\approx$~14~fF. The total Cooper-pair charging energy ($E_C = (2e)^2/2C_{tot}$) is $E_C/h$~=~0.78~GHz.  

The periodic structure of the metallic stripes, forms a phononic crystal, in which each stripe acts as an additional mass on the quartz surface. Thus, the SAW propagation velocity under the electrodes is, according to our estimates, about two percent less than on a free surface between them\cite{Morgan2007}. It provides the modulation of an effective refractive index $n$ between 1 and 1.02, forming a crystal lattice. The allowed modes in the crystal are plane Bloch waves, with a dispersion law strongly modified for the wavelengths close to the doubled crystal lattice period $P = a/2$ (half of the electric IDT period), corresponding to the edge of the first Brillouin zone. The wave group velocity is much smaller than the sound velocity and therefore the waves are effectively confined within the lattice. Hence, the phononic crystal with number of periods $N = 2 N_p$ acts similar to a resonator for the oscillations at the corresponding quantized frequencies. Acoustic waves from the transmission line are very weak and propagate into the bulk of the substrate. 
\\

{\bf Two-level system coupled to quasinormal modes} 

The IDT generates SAWs propagating in the longitudinal direction characterised by a wavevector $k_x$, and the transverse component characterised by a wavevector $k_y$ as shown in Fig.~\ref{Fig1}(b). Differently from the resonators, the waves are not reflected at the boundaries in the longitudinal direction but freely leaks out. The allowed modes therefore are quasinormal (QNM), rather than normal ones in resonators, characterized by $i$-th and $j$-th spacial modes in $x$ and $y$ directions. 

The Hamiltonian of our hybrid system can be written as 
\begin{equation}
\begin{aligned}
H  =- \frac{\hbar\omega_a}{2}\sigma_z & +\sum_{ij}^{} \hbar\omega_{ij}b_{ij}^\dagger b_{ij} \\
                                                               & + \sum_{ij}^{} \hbar g_{ij}(\sigma^+  b_{ij}^\dagger+\sigma^-b_{ij}), 
\end{aligned}
\label{Hamiltonian}
\end{equation}
where  $\hbar\omega_a$  is the energy splitting of the two-level system, $g_{ij}$ is the coupling to $ij$-th mode of the phononic crystal with frequency $\omega_{ij}$ and $b_{ij}$($b_{ij}^\dagger$) is the creation (annihilation) operator of the phonons in the quasinormal mode of the phononic crystal. We are interested in high quality modes. In such a case, the operators can be approximated by creation/annihilation operators of a phonon in a quantized system and Eq.~(\ref{Hamiltonian}) becomes identical to a Hamiltonian of a two-level system coupled to several quantized modes of harmonic oscillators.  

The artificial atom coupled to a phononic crystal interacts with the electromagnetic wave in the transmission line. The coupling capacitance $C_g$ between the atom and the line determines the interaction strength with the electromagnetic waves and noise in the line. For example, the relaxation rate $\Gamma_1$ of the system with the photon emission to the line is determined by coupling capacitance \cite{astafiev2010resonance,peng2016NatCom}.  
When an electromagnetic wave at frequency $\omega$ propagates through the line it drives the system  with the driving amplitude $\Omega$. The interaction Hamiltonian is $H_{int} = \hbar\Omega \sigma_x  \cos\omega t$.  
We measure the transmission coefficient of electromagnetic waves through the line $t=1-r$ (ratio of the transmitted to the applied wave amplitudes), where the reflection coefficient $r$ due to the scattered waves is  
\begin{equation}
r = i\frac{\Gamma_1}{\Omega}\langle \sigma^- \rangle.  
\label{Vsc}
\end{equation}
Eq.~(\ref{Vsc}) describes the dynamics of the scattered waves on the artificial atom, which are measured in the transmission spectroscopy. It contains information about interaction of the atom with the phononic modes. 
\\

{\bf Experimental Results} 

Our experiment is performed at a base temperature $T\approx$ 15 mK of a dilution refrigerator, so that the thermal fluctuations are well below the energy of surface acoustic phonons, which are in the gigahertz range of frequencies. We implement the measurement setup used for quantum optics experiments with superconducting artificial atoms, described in Refs.~[\citenum{wallraff2004strong,koch2007charge}]. The electromagnetic waves are transmitted from a vector network analyzer (VNA) through coaxial cables and a set of attenuators at different cooling stages, for suppressing room temperature blackbody radiation. Atom-wave interaction results in the scattering of the propagated through a transmission line waves, detected as a change in phase and amplitude of the transmitted signal close to the qubit resonance frequency. The method of measuring a spectrum of the artificial atom strongly coupled to a transmission line by scattering of propagated electromagnetic waves is described in Ref. [\citenum{astafiev2010resonance}]. The transmitted signal is then amplified by cryogenic and room-temperature amplifiers and measured by the VNA.

\begin{figure}
	\includegraphics[width=1.0\linewidth]{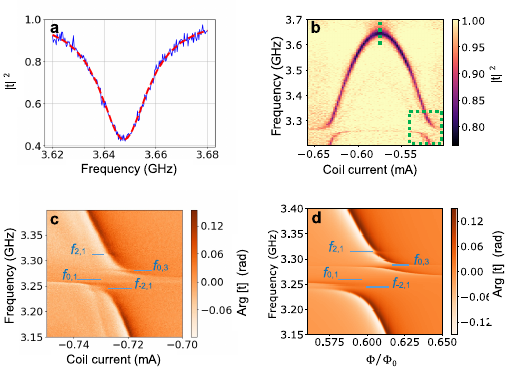}  
 	\caption{{\bf{Scattering spectroscopy.}} {\bf{a}}, An experimental curve (blue) of the transmission amplitude with a dip centered on the qubit transition frequency. It is fitted by a Lorentzian (red curve). {\bf{b}}, The qubit energy spectrum. The green vertical line shows the section where data for a plot (a) was measured. The green dashed rectangle represents a region of spectral line splittings shown in more details on a subplot (c).  {\bf{c}}, Spectral line splittings demonstrating interaction between the qubit and four QNMs of the phononic crystal at four frequencies.  {\bf{d}}, The simulated transmission phase colorplot obtained from simulations of the system. It reproduces the experimental anticrossings shown on (c).  
}
 	\label{Fig2}
  \end{figure}

A typical measurements of the transmission amplitude curve in vicinity of the qubit resonance is shown in Fig.~\ref{Fig2}(a). It has a dip because the atom scatters radiation back at resonant frequencies. The dip of more than 50\% indicates the strong coupling regime of our atom to the open transmission line. Fitting the linewidth by Loretzian curve, we extract the relaxation and dephasing rates: $\Gamma_1/2\pi$~=~8~MHz and $\Gamma_2/2\pi$~=~11~MHz\cite{astafiev2010resonance}.  By collecting such curves for various values of magnetic fields, controlled by the current through the coil, we find the energy splitting of the qubit. The spectral pattern is periodic in current with the period equal to one quantum of magnetic flux through the SQUID loop. The spectrum as a function of magnetic field around the maximal point in is shown in Fig.~\ref{Fig2}(b).

At frequency region close to $f_{ac}\approx$ 3.3 GHz, we find a set of four anticrossings as one can see in Fig~\ref{Fig2}(b). Some anticrossings demonstrate the strong coupling regime to the quasinormal modes of the phononic crystal and one of them is much stronger than others. These anticrossings are more clearly visible on a phase color plot shown in Fig.~ \ref{Fig2}(c). The mode frequencies are 3.244, 3.264, 3.283 and 3.313 GHz with splittings corresponding to the coupling strength $g$/2$\pi$ of  8, 39, 9 MHz. The highest mode was in the weak coupling regime and, therefore, the coupling is found with low accuracy.   

\begin{figure}
	\includegraphics[width=1\linewidth]{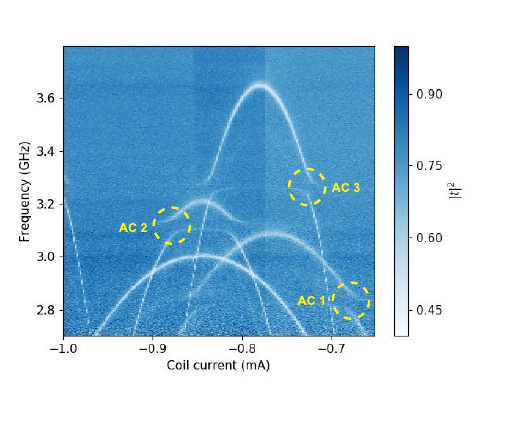}
 	\caption{{\bf{The spectrum of the control sample.}} Four qubits are designed with three different phononic crystal periods:   $a_1\approx 1.1~\mu m$, $a_2\approx 1.0~\mu m$, $a_3=a_4\approx 0.95~\mu m$. Three of these qubits demonstrate their interaction with QNMs at their predicted frequencies around 2.8 GHz (AC 1), 3.1 GHz (AC 2), and 3.3 GHz (AC 3). The fourth qubit spectrum is below its mechanical mode frequency, and, therefore, it does not have an anticrossing. 
	}
 	\label{Fig3}
  \end{figure}

To confirm that the feature comes from the lattice we have also fabricated and measured an additional control sample with four artificial atoms. The geometry of each atom is similar to the one  shown in Fig.~\ref{Fig1}. All the atoms are coupled to the transmission line and their scattering spectra are revealed on the same 2D-plot as it is seen in Fig.~\ref{Fig3}. Each atom is connected to its own phononic crystal with different periods: $a_1$ = 1.1 $\mu$m, $a_2$ = 1.0 $\mu$m, $a_3 = a_4$ = 0.95 $\mu$m. Therefore, we expect to observe up to four qubits with the strongest splittings at QNM corresponding to the frequencies $f_i \approx v/a_i$. Three qubits demonstrate a set of splittings on their spectra with major frequencies corresponding to calculated QNMs. One of the control qubits did not have anticrossing, since it is turned out that the qubit maximum frequency is lower than its acoustic mode of 3.26 GHz. 
\\

{\bf Data analysis and calculations } 

The first conclusion we can draw from the experimental data is that the wave is confined in $y$-direction, otherwise it will leak out within a short time. Taking into account dimensions of our device, we can find the diffraction angle to be about $\lambda$/$W\approx 0.1$, which will result in leaking out energy within less than 100 periods ($< 100~\mu$m) of our structure and the quality factor less than 100. The high quality factors ($\sim 1000$) in our experiment indicate that the diverging waves are confined in $y$-direction. In our geometry the lattice effective filling factor of metallic stripes (each much narrower than wavelength) of about $60\%$ results in the effective refractive index $n_{eff}\approx 1.012$ and the total reflection angle at the boundary of the lattice is $\alpha_c \approx \arcsin(1/n_{eff}) = 81^\circ$. We denote each quasinormal mode of our structure by two indexes $i,j$ meaning the mode orders in $x$ and $y$ component respectively. The allowed $k_y$ wavevectors are $k_y=\pi j/W$. With the condition $\alpha > \alpha_c$ for the incident angle $\alpha = \arctan(k_x/k_y)$, where $k_x \approx \pi/P = 2\pi/a$  we find that $j \leqslant 3$.

Among all quasinormal modes of the phononic crystal, the strongest coupling should be to the modes with frequencies close to $f_{ac}$ since the qubit is coupled to the acoustic modes through the stripes with period $a$. The atom-phononic field coupling strength for each $ij$-th can be calculated using the expression 
\begin{equation}
\hbar g_{ij}=\xi_0 C_{IDT}V_{ij},  
\label{g_general}
\end{equation}
where $\xi_0=\langle 0 |\partial H/\partial q|1\rangle$ is the qubit transition matrix element and $V_{ij}$ is the potential amplitude in $ij$-th mode induced in the IDT due to the mechanical zero-point fluctuations. For a transmon, $\xi_0$ can be found in the analytical form: $\xi_0 = (2E_c)^{3/4}E_J^{1/4}/e$. Importantly, $V_{ij}$ depends on the distribution of displacement amplitudes within the lattice, which for $ij$-th mode we represent as $z_{ij}(x,y)=z_0A_{ij}(x,y)$, where $z_0$ is the mean zero-point fluctuation displacement found as $z_0^2=\hbar/(2\rho W L v)$\cite{manenti2017circuit}  with mass density of quartz $\rho$. 

With our rectangular IDT, the field distribution can be simplified to the form $A_{ij}(x,y)=A_i(x)A_j(y)$ with the following normalization conditions $L^{-1} \int_{0}^{L} |A_i(x)|^2 dx$=1 and 
$W^{-1}\int_0^W|A_j(y)|^2 dy=1$. For the $y$-direction, the distribution can be easily found as $A_j(y)=\sqrt 2$ sin$(j \pi y/W)$ due to the boundary conditions $A_j(0) = 0$ and $A_j(W) = 0$. Now, we can find the potential induced on the IDT electrodes caused by the field in $x$-direction, which is an integral over area of corresponding metallic stripes. We introduce a function $p_e(x, y)$, describing the IDT electrode geometry \cite{Morgan2007}. It is approximated by a simple electrode step function defined as $\Pi(x, y) = 1$ for all $x$, corresponding to one electrode polarity and $\Pi(x, y) = -1$ for another polarity. $\Pi(x, y) = 0$ for all gaps between the electrodes. Taking into account that the function is independent of $y$ within $0 < y < W$, periodic in $x$-direction with the period $a$ and that the areas of different polarity electrodes are equal, we can write 
\begin{equation}
V_{ij}=\frac{e_{pz}}{\varepsilon}z_0\frac{\bigg|\int\limits_0^{L}\int\limits
_0^WA_{ij}(x,y)p_e(x,y)dydx\bigg|}{\frac{1}{2}\int\limits_0^{L}\int\limits
_0^W\big|p_e(x,y)\big|dydx},  
\label{Vij}
\end{equation}
where $e_{pz}/\varepsilon\approx1.6$ V/nm is the piezoelectric constant of quartz.\cite{Morgan2007}  After integration of Eq.~(\ref{Vij})  over y-direction we find that a simplified expression for the potential 
\begin{equation}
V_{ij}=\frac{e_{pz}}{\varepsilon}\frac{z_0 2 \sqrt 2}{j\pi}\frac{\bigg|\int\limits_0^{L}A_{i}(x)p_e(x)dx\bigg|}{\frac{1}{2}\int\limits_0^{L}\big|p_e(x)\big|dx},  
\label{Vij_2}
\end{equation}
where only odd $j$ are non-zero and can take only values $j = 1$ and $j = 3$. 

To calculate the spatial distribution of the field, frequencies $f_{i,j}$ and quality factors $Q_{i,j}$ for each mode we use an approach of quasinormal modes described in Refs. [\citenum{settimi2003quasinormal,severini2004second}]. Our calculations are based on the parameters of our device discussed above. We also take the structure metallization ratio of 0.65 from the device geometry. The approach is very much different from usual normal mode calculations without the phononic crystal with the zero or maximal field boundary conditions at the ends of the structure. The quasinormal modes are characterized by waves freely leaking out at the boundaries. The amplitudes of the leaking waves define the quality factors of the quasinormal modes. That is the total stored energy within the lattice over the leaking out power per period of oscillations in that particular mode gives us the quality factor. 

\begin{figure}
	\includegraphics[width=1\linewidth]{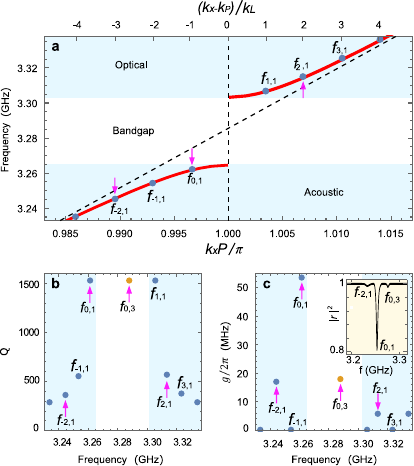} 
 	\caption{{\bf{The calculated parameters of QNMs.}} {\bf a}, The behavior of phonon dispersion curve (red) close to the first Brillouin edge. QNMs $f_{i,1}$ are depicted by blue points. The magenta arrows show the experimentally observed frequencies. {\bf {b,c}}, The quality factors (b) and coupling strength constant (c) for a set of QNMs close to a band gap (white rectangle). Quasinormal modes $f_{i,1}$  are depicted by blue points. An orange point correspond to $f_{0,3}$ mode. The experimental amplitude of a signal, reflected from the same geometry phononic crystal measured in a separate experiment is shown in the inset. Three dips correspond to the excitation of modes $f_{-2,1}=3.248$ MHz, $f_{0,1}=3.264$ MHz  and $f_{0,3}=3.283$ MHz, which have the highest coupling strength. The experimental $Q$-factors extracted from the widths of these dips are $Q_{2,1}$ = 496, $Q_{0,1}$ = 1040, $Q_{0,3}$ = 1100, which are in good agreement with calculated ones, shown on (b).
}
 	\label{Fig4}
  \end{figure}

The results of calculations are shown in Fig.~\ref{Fig4}(a-c). Figure~\ref{Fig4}(a) shows acoustic (lower) and optical (upper)
branches of the dispersion curves of the
phononic crystal near the edge of the first Brillouin zone calculated within Kronig-Penney model. Due to the finite size of our phononic crystal it has a discrete set of modes, depicted by separate blue dots on the continuous dispersion curves. Figure~\ref{Fig4}(b) shows frequencies and quality factors of several modes close to $f_{ac}$. Blue dots correspond to a set of quasinormal modes $f_{i,1}$. Orange dots correspond to $f_{0,3}$ mode. Three modes $f_{0,1}$, $ f_{1,1}$ and $f_{0,3}$, which are closest to the band gap, have the highest quality factors of about 1500. For modes far from the gap, the quality factors rapidly decrease, which is in good agreement with the width of experimentally observed resonance dips (black curve in the inset) of the signal reflected from the crystal.  

\begin{figure}
	\includegraphics[width=1\linewidth]{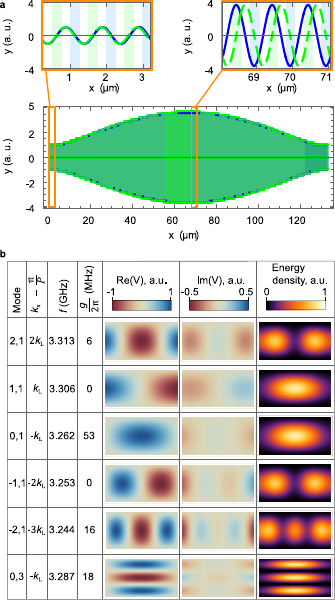} 
 	\caption{{\bf{The field distribution of QNMs.}} {\bf{a}}, The spatial dependence of the induced potential $V_i(x)$ on the pairs of electrodes due to the quasinormal mode $f_{0,1}$ (blue) and  $f_{1,1}$ (green). The insets show field details in respect to the electrodes of the IDT. Blue and green colors indicate electrodes of opposite electric polarity. {\bf{b}} The colormaps for real (left column) and imaginary (central column) part of the complex potential amplitudes, calculated as a field difference on pairs of electrodes for several different modes. The right most plots show energy distribution in acoustic waves.    
}
 	\label{Fig5}
  \end{figure}

Figure~\ref{Fig4}(c) shows the calculated values of the coupling constants $g$ for modes near the band gap. There is a significant relative difference in these constants. It is a direct consequence of the differences of the modes field distribution functions  $A_{ij}(x)$, which affect the value of integral in Eqs.~(\ref{Vij},\ref{Vij_2}). In particular, for the mode $f_{0,1}$  the coupling is large, while for an adjacent odd mode $f_{1,1}$ it is zero. The field distribution for these two modes is shown in Fig.~\ref{Fig5}(a). In both cases it is symmetric and has maximum in the centre. The fundamental difference is that the field for the $f_{0,1}$  mode has antinodes on the electrodes, while the field for the $f_{1,1}$ mode has its nodes on the electrodes. Odd modes are decoupled from the atom. Frequencies $f_{0,1}$, $f_{2,1}$, $f_{-2,1}$ and $f_{0,3}$ perfectly match to the calculated ones in Fig.~\ref{Fig4}(a) and marked by magenta arrows. 

Figure~\ref{Fig5}(b) shows spatial distribution of potentials induced by mechanical waves under the IDT stripes for several modes near the Brillouin zone, namely real and imaginary parts of potential differences at each pair of IDT electrodes $V = \frac{e_{pz}}{\varepsilon}[A_{ij}(x_{2n}, y) - A_{ij}(x_{2n+1}, y)]$, where $x_n =  nP = na/2$ with integer $n \geq 0$. The imaginary part of the field is related to the decay of the waves. The plots on the right hand side show energy distribution in the acoustic field. The field spacial distribution in $x$-direction depends on $k_x$ and take values $k_P + i k_L$ for $i>0$ and $k_P + (i-1) k_L$ for $i\leq0$, where $k_P = \pi/P$ and $k_L = \pi/L$. Note that the mode $k_x = k_P$ does not exist. The coupling depends on the symmetry of the modes in their space distribution.  From the induced potential distribution of Fig.~\ref{Fig5}(b), it is obvious that antisymmetric field distributions (odd modes) result in averaged zero field and are not coupled to the atom. Oppositely, even modes give non-zero coupling but the coupling strength is rapidly decreasing with the mode number. Modes (1,~1) and (0,~1), exemplified in Fig.~\ref{Fig5}(a) have the same energy distribution (acoustic wave amplitude-square) but excite different potentials under the stripes and therefore lead to different coupling strengths. 

Quality factors of different size structures have been studied on several samples. As expected, quality factor increases with increasing the number of stripes because it takes a longer time for the wave to reach the boundaries of the structures. This is also supported by simulations. However, increasing the number of stripes weakly affects the coupling strengths as the field amplitudes scale inversely proportionally to square-root of the system length. 

Finally, we find the correspondence between the experimentally measured (Fig.~\ref{Fig2}(c)) and simulated modes having frequencies $f_{-2,1}$~=~3.244~GHz, $f_{0,1}$~=~3.262~GHz,  $f_{0,3}$~=~3.287~GHz, $f_{+2,1}$~=~3.313~GHz. With Eq.~(\ref{g_general}) to calculate coupling strengths and get $g_{-2,1}/2\pi\approx 17$~MHz, $g_{0,1}/2\pi\approx 53$~MHz, $g_{0,3}/2\pi\approx 18$~MHz and $g_{2,1}/2\pi\approx 6$~MHz. Calculated $g_{i,j}$ are in a good agreement with the experiment. Thus, the approach of quasinormal modes predicts correctly the field distributions  $A_{i,j} (x)$. On the other hand, all the couplings are slightly higher, than the experimental ones. We suppose that this is a consequence made on the approximations of the geometry factor $p_e(x,y)$. Due to the field distribution, coupling to the mode (0,1) is several times stronger comparing to others. This can be helpful for realisation of effects of quantum optics and quantum acoustics, in which the artificial atom is mainly coupled to a single mode.   

In order to simulate the signal shown in Fig.~\ref{Fig2}(d) we solve the Master Equation and find an expectation value of the atomic state annihilation operator $\langle\sigma^-\rangle$, since it defines the amplitude of the electromagnetic wave scattered from the atom according to Eq.~(\ref{Vsc}). 
The Hamiltonian and decay operators are built using calculated above QNMs coupling strengths $g_{i,j}$, frequencies $f_{i,j}$  and quality factors $Q_{i,j}$ (see Fig.~\ref{Fig4}(b)). The Fig.~\ref{Fig2}(d) shows phase for the simulated transmission, which is in a good agreement with the experimental data. 

In conclusion, we experimentally demonstrate the interaction between a qubit and a surface acoustic wave phononic crystal formed by a periodic metallic structure on a surface of quartz. In our circuit, the modes of the phononic crystal are found by characterisng the scattering of electromagnetic waves on a two-level artificial atom strongly coupled to the crystal. We have found interaction of the atom with four quasinormal modes of the crystal. The quasinormal modes and their properties are compared from independent calculations. Our geometry is simple and robust in fabrication. It is more compact than setups involving mirrors and other elements. Our results contribute to fundamental quantum acoustics and can be useful for developing devices of quantum acoustics. 

{\bf Data availability.}
Relevant data is available from A.N.B. upon request.

 {\bf Author contributions.}
O.V.A. planned and designed the experiment, A.N.B., D.D.K. and R.S.Sh. fabricated the sample. A.Yu.D, A.N.B., D.D.K and R.S.Sh built the set-up for measurements and measured the raw data, A.V.S., S.V.S., A.N.B and O.V.A. made calculations, analysed and processed the data and wrote the manuscript.  
 
 {\bf{Acknowledgements.}}
We acknowledge Russian Science Foundation (Grant No. 16-12-00070) for supporting the work. This work was performed using technological equipment of MIPT Shared Facilities Center.

 {\bf Competing interests.}
The authors declare no competing financial interests.



\bibliography{Main_Resubmit}

\end{document}